\documentclass[10pt,a4paper]{siamltex}
 \usepackage{graphicx}   
\usepackage[latin1]{inputenc}
\usepackage{amsmath}

\usepackage{amscd,amssymb}

\title{Crossing paths in  2D Random Walks}
\author{Marc Artzrouni \\
Department of Mathematics \\
University of Pau; 64000 PAU; FRANCE}
\begin{document}
\maketitle
\begin{abstract}
We investigate crossing path probabilities  for two agents that move randomly in a bounded region of the plane or on a sphere (denoted $R$).  At each discrete time-step the agents  move, independently,  fixed distances $d_1$ and $d_2$  at angles that are  uniformly  distributed  in $(0,2\pi)$.  If $R$ is large enough  and the initial positions of the agents are uniformly  distributed  in $R$, then the probability  of paths crossing  at the first time-step is close to $ 2d_1d_2/(\pi A[R])$, where $A[R]$ is the area of $R$.  Simulations suggest that the long-run rate at which paths cross is also  close to $2d_1d_2/(\pi A[R])$  (despite marked departures from uniformity and independence conditions needed for such a conclusion).  
\end{abstract}

 \section{Introduction}\label{intro}
 Random walks  have been studied  in  abstract settings such as   integer lattices $\mathbb{Z}^d$  or  Riemannian manifolds (\cite{Law96}, \cite{Rob60},   \cite{Telc06}). 
In applied settings  there are many  spatially explicit individual-based  models (IBMs) in which  the behavior of the system is determined by the meeting  of  randomly moving agents. The transmission of a pathogenic agent,   the spread of a rumor, or the sharing of some property when randomly moving particles meet are examples that come to mind  in biology, sociology,  or physics (\cite{Hol06}, \cite{Stef06},  \cite{Per96}, \cite{Par05}, \cite{Ber93}).  In many of these models the movement of agents is conceptualized as discrete  transitions between square  or hexagonal cells (\cite{Hol06}). However, such a stylized representation of individual movements  may not always be entirely realistic. 

Although IBMs are powerful tools for the description of complex systems, they suffer from a shortage of  analytical results.  For example, if a susceptible and an infective agent move randomly in some bounded space, what is the probability of them meeting, and hence of the transmission of the infection? What is the average time until the meeting takes place? 

In the present paper we begin  to answer these questions by considering a  random walk in a bounded region of the  plane or on the sphere, which we denote by $R$. The model evolves in discrete time.  At each time-step an agent leaves its current position at a uniformly distributed angle in the $(0,2\pi)$ interval. On the plane the agent  moves a fixed distance $d$  in a straight line.  On a sphere the agent moves a fixed distance $d$ on a geodesic.

Here we will consider two such agents who move  different distances $d_1$ and $d_2$ at each time-step. We assume that the agents' initial positions are uniformly  distributed in $R$. The paper's central results concern the probability that the paths of the two agents cross during the first time-step.  If  $R$ is either a sufficiently large  bounded region of the plane or a sphere, this "first-step" probability of intersection   is close to $2d_1d_2/(\pi A[R])$ where $A[R]$ is the area of $R$.   

In applied settings we are often interested in the  \textit{long-run average rate} at which the paths cross.  In order to extend results on the "first-step" probability of intersection and apply the law of large numbers  we would need the following assumptions: 

\begin{itemize}
\item The positions of the two agents are uniformly distributed \textit{at every time step} (which is the case on the sphere but not on the plane because of reflection problems at the boundary of $R$),
\item The crossing-path events are independent over time (which is the case in neither setting  because of the   strong spatial dependence at consecutive time-steps). 
\end{itemize}
 
Numerous simulations have shown  that despite marked departures from these assumptions the long-run rate at which the paths cross is  also close to   $2d_1d_2/(\pi A[R])$. 
% \linebreak[4] 

Section \ref{mod}   contains the results both for the plane and the sphere.  
Section \ref{sim}   is devoted to the numerical simulations. Extensions are discussed in Section \ref{disc}.   Three technical appendices can be found in  Section \ref{app}. 

\begin{figure}
\includegraphics[scale=0.8]{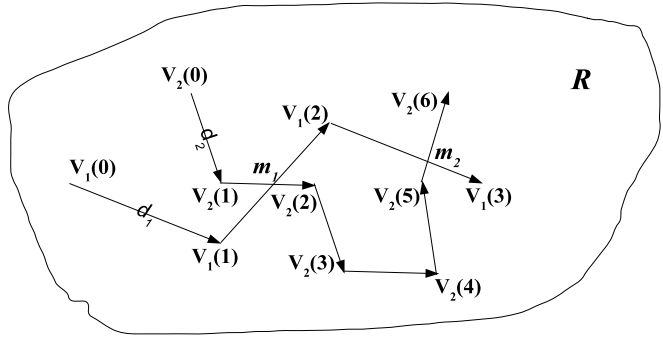} \\
\caption{Example of two trajectories ${V_1(k)}$ and ${V_2(k)}$ with paths that crossed synchronously  at $m_1$  for $k=2$.
(We are not interested in the asynchronous  crossing  at $m_2$ which  took place at different times for the two agents).} 
\label{labelfig0}
 \end{figure}

 \section{Model}\label{mod}
 \subsection{Geometric description in the plane and on the sphere}\label{geom}
 A bounded region of the plane is the most natural setting for agents moving in a 2D environment. However we then need  to specify how agents are reflected when  they hit the boundary of the region.  There is no such problem on a sphere.   In what follows $R$ is either a bounded region of the plane or a sphere. 

The initial positions  ${V_1(0)}$ and ${V_2(0)}$ of the two agents are assumed uniformly distributed in $R$.  At each time-step the two  agents  move  distances  $d_1$ and $d_2$ (which are fixed positive parameters) in a straight line (or along a geodesic on a sphere). They depart at random angles $\alpha_1$ and $\alpha_2$ that are uniformly and independently  distributed over $(0,2\pi)$.   The endpoints after the $k-th$  time-step are  $V_1(k)$ and $V_2(k)$  (Figure \ref{labelfig0}). 

The definition of a meeting in such a model  is tricky because  the probability of the two agents being in exactly the same position at any given period is 0.   There are  however different ways of approximating such a meeting. 

One could say that the agents meet if the distance between two points ${V_1(k)}$ and ${V_2(k)}$ is less than some $\epsilon$. In such a definition results would depend on $\epsilon$, which is undesirable.  
For this reason  we choose to define a meeting during the $k-th$ time-step  when paths cross between  $k$ and $k+1$.  This means that  the segments (or the "geodesic arcs")   $V_1(k)V_1(k+1)$ and $V_2(k)V_2(k+1)$ intersect.  We recognize that ${V_1(k)}$   and ${V_2(k)}$ can then be close without the paths crossing, but at least the definition does not depend on some arbitrary $\epsilon$.  

The point $m_1$ in Figure \ref{labelfig0} is an example of such a synchronous crossing of paths.  Of course  the agents are not at $m_1$ at the same time.  In the figure the paths cross asynchronously at $m_2$.

Much will depend on whether  ${V_1(k)}$ and ${V_2(k)}$  are uniformly distributed on $R$ for every $k$.  This will be the case if $R$ is a sphere because 
${V_1(0)}$ and ${V_2(0)}$  are themselves uniformly distributed. If on the other hand $R$
 is a bounded region of the plane, then the uniformity in the distributions of  ${V_1(k)}$ and ${V_2(k)}$ is compromised  for $k>0$   by  the vexing problem of
 the behavior of the agents when they hit  the boundary of $R$.  

For this reason we focus on the probability of paths crossing at  the first time-step only.  We simplify notations by letting 
 $V_1\overset{def.}{ =} (x_1, y_1)$ and $V_2\overset{def.}{ =} (x_2, y_2)$ be the initial positions of the two agents.  The corresponding  endpoints are denoted  $W_1$ and  $W_2$.   

We next proceed with calculations when  $R$ is a bounded region of the plane.

\subsection{Crossing-path probability  in a bounded region of the plane}\label{prob}

In the plane the endpoints   $W_1$ and $W_2$ are  

\begin{equation}\label{eqw1}
W_1\overset{def.}{ =} V_1+ d_1( cos(\alpha_1),  sin(\alpha_1))  
 \end{equation}
\begin{equation}\label{eqw2}
 W_2 \overset{def.}{ =} V_2+ d_2( cos(\alpha_2),  sin(\alpha_2))   
 \end{equation}
where $\alpha_1$ and $\alpha_2$  are polar angles  that are uniformly distributed in $(0,2\pi)$.  

We will finesse the reflection problem at the boundary of  $R$ by  considering the possibility 
of intersection on the basis of  Eqs.   \eqref{eqw1} -  \eqref{eqw2} even if $W_1$ or  $W_2$ is  outside $R$.  In such a case we
 examine first  whether the intersection has occurred. We then  move, in some unspecified manner, 
 the wayward point(s) back inside $R$. 

We now define  the \textit{ feasible  domain} $ FD(V_1,\alpha_1)$ as  the set of points $V$  that are within $d_2$ of the segment $V_1W_1$:
\begin{equation}\label{FD}
  FD(V_1, \alpha_1) \overset{def.}{ =} \left\{  V=(x, y)  \mbox{ such that }  d(V,  (V_1W_1)) \le d_2  \right\}
       \end{equation}
where $d(V,  (V_1W_1))$ is the distance between a point $V$ and the segment $V_1W_1$. 

\begin{figure}
\includegraphics{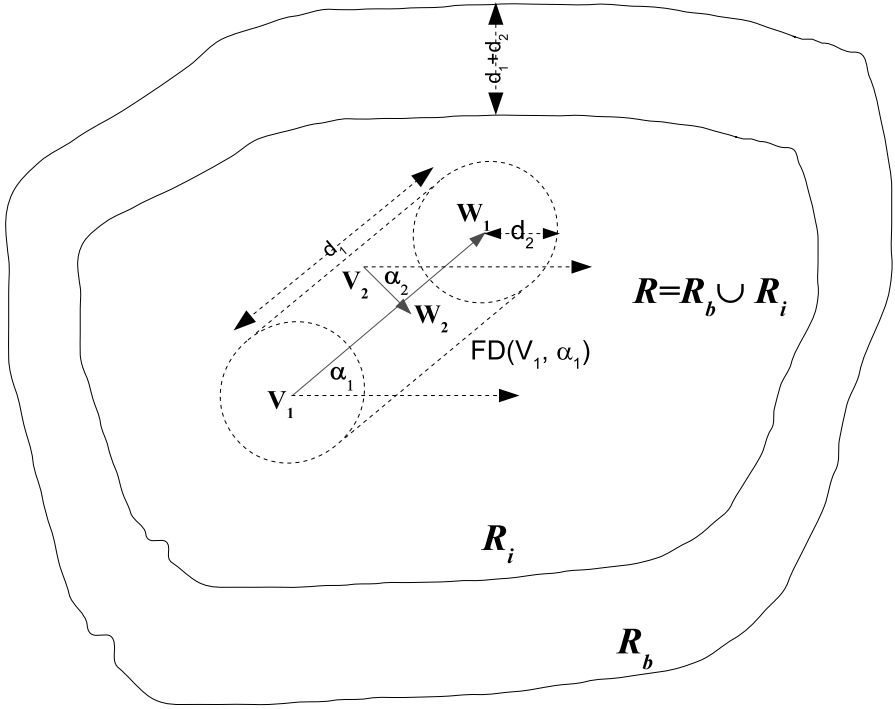} \\
 \caption{Example of two intersecting segments $V_1W_1$ and $V_2W_2$   with feasible domain $FD(V_1, \alpha_1)$. Inner and  border regions  
$R_i$ and  $R_b$ are  at distances to  the outside world that are larger than and less than  the sum $d_1+d_2$. (The distance $d_1+d_2$ between the two  boundaries is not to scale).}
\label{labelfig1}
 \end{figure}

%%%%%%%%% SEE BELOW TO FORCE line break just before math expression that would make line go over - 
% number (between 1 and 4) is a measure of how insistent  I am about the line break - don't ask - 
%%%%%       \linebreak[4]     THIS is for line break if a math expression makes the line go over *************************

The point  $V_2$ must be in $FD(V_1, \alpha_1)$ in  order for the segments $V_1W_1$ and  $V_2W_2$ to  intersect - although the condition is not sufficient. 
Figure \ref{labelfig1} depicts a segment  $V_1W_1$ and the corresponding feasible domain $FD(V_1, \alpha_1)$. This domain   is bounded by  dotted lines (a rectangle with sides $d_1$ and $2d_2$ 
with  a half circle of radius $d_2$  at each end of the rectangle). In the figure a point $V_2$ is in the feasible domain and the  two segments
 $V_1W_1$ and $V_2W_2$  intersect. Whether there is an intersection depends on the angle $\alpha_2$. 

The region $R$ is partitioned into an inner  region $R_i$ and a border region $R_b$ characterized by a distance to the outside world that is 
either larger  (for  $R_i$) or smaller (for  $R_b$)   than $d_1+d_2$.   (The point is that $FD(V_1, \alpha_1)$ is entirely in $R$ if  $V_1\in R_i$).
We  define$A[S]$ as the area of  a bounded subset  $S$ of $\mathbb{R}^2$.

%  If  $V_1$ falls in  $R_i$ then the corresponding feasible domain $FD(V_1, \alpha_1)$ is entirely in $R$ (regardless of $\alpha_1$). The point  $V_2$, chosen uniformly in R,  falls then  in $FD(V_1, \alpha_1)$ with a probability equal to $A[FD(V_1, \alpha_1)]/A[R]$. 
%
% If  $V_1$ falls in  $R_b$ then at least for some values of $\alpha_1$ there will be a  part of $FD(V_1, \alpha_1)$ that is outside  $R$. In this case  $V_2$ falls  in $FD(V_1, \alpha_1)$
% with a probability equal to $A[FD(V_1, \alpha_1)\cap R]/A[R]$ that is smaller than  $A[FD(V_1, \alpha_1)]/A[R]$.    
% 
With  $V_1=(x_1, y_1) $ and  $V_2=(x_2, y_2)$ 
uniformly and independently distributed on $R$, we will now calculate  the  probability that   $V_1W_1$ and  $V_2W_2$  intersect. 
This first-step probability of intersection  depends only on $d_1, d_2$ and $R$, and is noted $P_p(d_1,d_2,R)$. (The subscript $p$ indicates a probability in the plane).

We first need  the probability $G_p(x_1,y_1, \alpha_1 )$  of the intersection conditionally on $V_1=(x_1,y_1)$ and  $\alpha_1 $; 
$G_p(x_1,y_1, \alpha_1 )$ is the probability (denoted by $p_1$ below)   that  $V_2$ falls in $FD((x_1,y_1), \alpha_1)\bigcap R$   multiplied  by the probability (denoted by $p_2$ below) 
of the segments intersecting conditionally on  $V_2$ falling in $FD((x_1,y_1), \alpha_1)\bigcap R$.    (In the $FD$ function we have replaced $V_1$ by its coordinates $(x_1, y_1)$).  

Given  the uniformity assumption on $V_2$, the probability $p_1$ is then 

\begin{equation}
p_1=\dfrac{A[FD((x_1,y_1),\alpha_1)\bigcap R]}{A[R]}.
\end{equation}

 In order to calculate $p_2$ we need to define the probability  $f_p(x_1,y_1,\alpha_1,x_2,y_2)$ that  $ V_2W_2$ intersects   $V_1W_1$, conditionally 
on  $V_1=(x_1,y_1), V_2=(x_2,y_2)$ and  the angle $\alpha_1$.  (This probability, derived in  Appendix A,  is obtained by calculating  the magnitude of the angle
  $\beta$ within which  $\alpha_2$ must fall for the intersection to occur, and then dividing by $2 \pi$).  The conditional probability $p_2$ of the intersection is then  

\begin{equation}
p_2=\dfrac{\iint_{( x_2,  y_2) \in FD((x_1,y_1),\alpha_1)\bigcap R} f_p(x_1,y_1,\alpha_1 , x_2,y_2   )dx_2dy_2 }{A[FD( (x_1,y_1) ,\alpha_1)\bigcap R]}. 
\end{equation}

The probability   $G_p(x_1,y_1, \alpha_1)$   is now  the product $p_1p_2$ which  simplifies to
\begin{equation} \label{G10}
G_p(x_1,y_1, \alpha_1 )= \dfrac{\iint_{(x_2,  y_2) \in FD( (x_1,y_1),\alpha_1)\bigcap R} f_p(x_1,y_1,\alpha_1,x_2,y_2)dx_2dy_2 }{A[R]}.
\end{equation} 

When  $V_1=(x_1,y_1)$ is in $R_i$ (i.e. the feasible domain is entirely in $R$) then the double integral on the right-hand side of 
Eq. \eqref{G10} is independent of  $V_1$ and is noted $I_p(d_1, d_2)$, i.e. 

\begin{equation}\label{G190}
I_p(d_1, d_2)  \overset{def.}{ =} \iint_{(x_2,  y_2) \in FD( (x_1,y_1)   ,\alpha_1)} f_p(x_1,y_1,\alpha_1,x_2,y_2)dx_2dy_2.
\end{equation}
The quantity   $I_p(d_1, d_2)$ is  an upper bound for  the double integral     in Eq. \eqref{G10}   when
  $V_1=(x_1,y_1)$ is in the border region $R_b$ (because the integration is then over an area smaller than the feasible domain $FD(x_1,y_1)$).  

We now have the following result on $I_p(d_1, d_2)$, $G_p(x_1,y_1, \alpha_1 )$, and  $P_p(d_1,d_2,R)$.
\begin{proposition}\label{PROP1} We have  
\begin{equation}\label{G19}
 I_p(d_1, d_2) = 2d_1d_2/\pi    
\end{equation}
and when  $V_1=(x_1,y_1)$ is in $R_i$, 
\begin{equation}\label{GGc}
   G_p(x_1,y_1, \alpha_1 )      =  \dfrac{2d_1d_2}{\pi A[R]}.
\end{equation}
When  $V_1=(x_1,y_1)$ is in $R_b$  then
\begin{equation}\label{GGa}
G_p(x_1,y_1, \alpha_1 ) \le  \dfrac{ 2d_1d_2}{\pi A[R]}         .
\end{equation}
The first-step probability of intersection $P_p(d_1,d_2,R)$ satisfies 
%  \eqref{P2100} \eqref{P2200}
\begin{equation}  \label{P2100}
    p_{\ell o}    \overset{def.}{ =}  \dfrac{2d_1d_2A[R_i]}{\pi A[R]^2}   \le    P_p(d_1,d_2,R)  \le    p_{hi}    \overset{def.}{ =}  \dfrac{2d_1d_2}{\pi A[R]}
\end{equation}
which leads to the mid-point approximation
\begin{equation}  \label{P2200}
 P_p(d_1,d_2,R)   \approxeq    p^*  \overset{def.}{ =} d_1d_2 \dfrac{A[R_i]+A[R]}{\pi A[R]^2}. 
\end{equation}

The absolute value of the  maximum percentage error (AVMPE) made with the approximation of   (\ref{P2200})  is  
\begin{equation}  \label{P23}
 AVMPE =  100 \dfrac{A[R]-A[R_i]}{A[R]+A[R_i]}.
\end{equation}

\end{proposition}
\begin{proof}See Appendix A.  
\end{proof}

\textit{Remark.}  The approximation of Eq. \eqref{P2200} is of interest and the error of Eq. \eqref{P23} is small only if $A[R]$ is large enough in the sense that  $A[R_i]$ is \textit{relatively close} to $A[R]$.
 This means there is a large subset of $R$ within which the feasible domain $FD(V_1,\alpha_1)$ is entirely in $R$. Suppose for example 
 that $R$ is a circle of radius  $r>d_1+d_2$, then 
\begin{equation}  \label{P24}
 AVMPE =  100 \dfrac{ 1- \left(  1- \frac{d_1+d_2}{r} \right) ^2} {  1+  \left(  1- \frac{d_1+d_2}{r} \right) ^2       } 
\end{equation}
which is approximately $100(d_1+d_2)/r$ when $d_1+d_2$ is much smaller than  $r$. 
 Therefore if   $d_1+d_2$ is one percent of the radius then the maximum  error made with the estimate of Eq. \eqref{P2200} is also
 approximately  one percent. 

We next turn our attention to the situation in which   $R$ is a sphere.
%In the Appendix we show that if $(x_1,y_1)\in R_i$ (i.e. the feasible domain is entirely in R) then the double integral on the right-hand side of Eq. \eqref{G10}  is $2d_1d_2/ \pi$ and 
%therefore   $G(x_1, y_1, \alpha_1)=2d_1d_2 /(A[R] \pi)$.  Equation   (\ref{P1})  is then
 
%When modeling for example  the spread of an infectious agent that depends on randomly moving agents, we are really interested in the  long-run rate at which the paths cross.   \textit{If}  $V_1(k)$ and $V_2(k)$ could be considered uniformly distributed at each time-step $k$, then the probability of the paths crossing  would be $P(d_1,d_2,R)$ at each time-step.  \textit{If} in addition the law of large numbers were applicable (which it is not)   then the long-run meeting rate would also be approximately  $P(d_1,d_2,R)$.   These issues  will be explored through simulations in Section \ref{sim}.
%% for conclusion: add alpha = prob of transmission givne a metting - as on obvious generalization - 

\subsection{Crossing-path probability on a sphere}\label{sph}
When  the domain $R$ is a sphere of radius $\rho$ we use the spherical system of coordinates  where a point $V$   is defined by the triplet $(\rho,\theta, \phi)$ of radial, azimuthal, and zenithal coordinates. 

Given   initial points  $V_k \overset{def.}{ = } (\rho, \theta_k, \phi_k), (k=1, 2)$   each  endpoint $W_k$ is on the circle  at a geodesic distance $d_k$ from $V_k$. The position of $W_k$ on the circle is determined by an angle $\alpha_k$ uniformly distributed in $(0, 2\pi)$.  See Appendix B for the exact derivation of the endpoints $W_k$. 

% $\overrightarrow{\tau_2}$ 

%Each agent departs its  initial position  $V_k \overset{def.}{ = } (\rho, \theta_k, \phi_k), (k=1, 2)$ and moves a  geodesic distance $d_k$   in the direction of  vector $\overrightarrow{\tau_k}$ that is tangent to the sphere at $V_k$ and  at a uniformly distributed angle $\alpha_k$   on the tangent planes to the sphere at $V_k$ 

We let $P_s(d_1, d_2, R)$   be the probability  that the arcs $V_1W_1$ and  $V_2W_2$ intersect.  Because the  initial points  $V_k$ are uniformly distributed, the endpoints will also  be  uniformly distributed on the sphere. Therefore $P_s(d_1, d_2, R)$  is the probability of paths crossing at every time-step.

 In order to calculate  $P_s(d_1, d_2, R)$ we  proceed as before except that there is no border area and the double integrals are  calculated in spherical coordinates.  The feasible domain is denoted $FD(\theta_1,\phi_1)$ and is defined with geodesic distances.  The differential area element is now $\rho ^2sin(\phi)d\theta d\phi$. 

We let $f_s(\theta_1,\phi_1,\alpha_1,\theta_2,   \phi_2)$ be the probability of intersection conditionally  on the arc $V_1W_1$ (defined by  $ \theta_1, \phi_1$ and $\alpha_1$) and on $V_2$ (defined by $\theta_2$ and $\phi_2$).   We also   define the double integral 

\begin{equation} \label{G10se}
I_s(d_1, d_2,\rho)     \overset{def.}{ = }  \iint_{(\theta_2,  \phi_2) \in FD( (\theta_1,\phi_1)   ,\alpha_1)} f_s(\theta_1,\phi_1,\alpha_1,\theta_2,   \phi_2)\rho^2sin(\phi_2)d\theta_2d\phi_2.  
\end{equation} 

The   probability   of the intersection conditionally on 
$V_1=(\rho, \theta_1,\phi_1)$ and  $\alpha_1$ is independent of  $\rho, \theta_1,\phi_1$ and is denoted   $G_s(d_1, d_2, \rho)$. We then have
\begin{equation} \label{G10s}
 G_s(d_1, d_2, \rho)= \dfrac{ I_s(d_1, d_2,\rho) }{A[R]}
\end{equation} 
where the area $A[R]$ of the sphere appearing in the denominator is now $4\pi\rho^2$. 

The probability of paths crossing at each time-step   is then 
\begin{equation*}
P_s(d_1,d_2,R) = \dfrac{\iiint_{(\theta_1, \phi_1) \in R, \alpha_1 \in (0,2\pi )}   G_s(\theta_1,\phi_1, \alpha_1)    \rho^2sin(\phi_1)    d\theta_1d\phi_1d \alpha_1}   {2\pi A[R]} 
\end{equation*} 
\begin{equation}\label{G10u}
% \dfrac{I_s(d_1, d_2,\rho) \iiint_{(\theta_1, \phi_1) \in R, \alpha_1 \in (0,2\pi )}   \rho^2sin(\phi_1)    d\theta_1d\phi_1d \alpha_1}   {2\pi A[R]^2} = 
= \dfrac{ I_s(d_1, d_2,\rho) }{A[R]}  = \dfrac{ I_s(d_1, d_2,\rho) }{ 4\pi\rho^2 }.
\end{equation} 
When $\rho\rightarrow\infty$ the integral $I_s(d_1,d_2,\rho)$  on the sphere approaches the corresponding integral  $ I_p(d_1, d_2) = 2d_1d_2/\pi$ on the plane (Eq. \eqref{G19}). 

In the next proposition we show numerically that $I_s(d_1,d_2,\rho)$ is extremely close to $2d_1d_2/\pi$  even when $\rho$ is not particularly large compared to $d_1$ and $d_2$. We will just assume that 
\begin{equation}\label{G10uz}
\dfrac{d_2}{\rho} < \dfrac{\pi}{2}, \dfrac{d_1}{2\rho} + \dfrac{d_2}{\rho}  < \pi
\end{equation}
which insures that the feasible domain does not "wrap around" the sphere. 
  \pagebreak
\begin{proposition}\label{PROP2}
With   (\ref{G10uz})  we have  
\begin{equation}\label{G10tt}
 I_s(d_1, d_2,\rho)      \approxeq   2d_1d_2/\pi
\end{equation}
and the probability of paths crossing at each time-step is 
\begin{equation}\label{GGb}
 P_s(d_1,d_2,R) \approxeq   \dfrac{2d_1d_2}{\pi A[R]} =  \dfrac{d_1d_2}{2\pi^2\rho^2}. 
\end{equation}
The  differences between both sides of  (\ref{G10tt}) and   (\ref{GGb})   are so small that they are within the margins of error when calculating $I_s(d_1, d_2,\rho)$ (and
$ P_s(d_1,d_2,R)$)  numerically. 
\end{proposition}
\begin{proof}See Appendix C.  
\end{proof}

The next section is devoted to simulations aimed at assessing the quality of the approximations derived above. 

\section{Simulations}\label{sim}
We consider a  region $R$ that is a circle of radius $r=10$. At each time-step the two agents move distances $d_1=1$ and $d_2=0.7$ respectively.
If an agent moves outside the circle we first check whether paths have crossed. We then move the wayward agent  to a 
point diametrically opposed to its current position, at a distance inside the circle equal to the distance  between the circle and the current position. This algorithm could be considered  a 2D version  
of  the wrapping around that takes place on a sphere.   The goal is to try to keep the distribution of the agents as uniform as possible.  We need this  in order for the crossing-path probability to 
remain  as close as possible to the "first-step" probability derived under the assumption that initial positions are uniformly distributed. 
% calc below in MAINPROOFCheck3DOCT07.mcd  ; done Nov 30, 2007

The bounds of  \eqref{P2100} and the approximation of    \eqref{P2200} are now used to calculate  the low,  high and  mid-point approximations $p_{\ell o}, p_{hi}, p^*$ 
  for the first-step probability of intersection  $P_s(d_1,d_2,\rho)$:      
\begin{equation}  \label{P2800}
 p_{\ell o}     = \dfrac{2d_1d_2A[R_i]}{\pi A[R]^2}= 0.0009772,   \bigskip  p_{hi} = \dfrac{2d_1d_2}{\pi A[R]}=0.001418,
\end{equation}
\begin{equation}  \label{P2801}
 p^*= d_1d_2 \dfrac{A[R_i] + A[R_b]}{\pi A[R]^2}   =0.001198
\end{equation}
which  translate into a maximum error $AVMPE$ on  $p^*$ of 18.42\% (Eq. \eqref{P24}). This error is relatively large because the sum $d_1+d_2$ is 1.7, which is not particularly small compared to the radius 10 of the circle. 
 
We let $F(k)$ be the  random variable equal to the average crossing path   frequency over the first $k$ time-steps.  \textit{If}  the probability of paths crossing  at every time-step were   $p_{\ell o}$ (or $p_{hi}$)  and \textit{if} paths crossing  were independent events (which they are not)  then  the Central Limit Theorem would insure  that  for large $k$ the corresponding  "hypothetical frequency"  $F_h(k)$  would be   approximately normally distributed with mean  $p_{\ell o}$ (or $p_{hi}$)  and standard deviation  $\sqrt{p_{\ell o}(1-p_{\ell o})/k}$ (or $\sqrt{p_{hi}(1-p_{hi})/k}$).  

\begin{figure}[h]
\begin{center}
\includegraphics[scale=0.7] {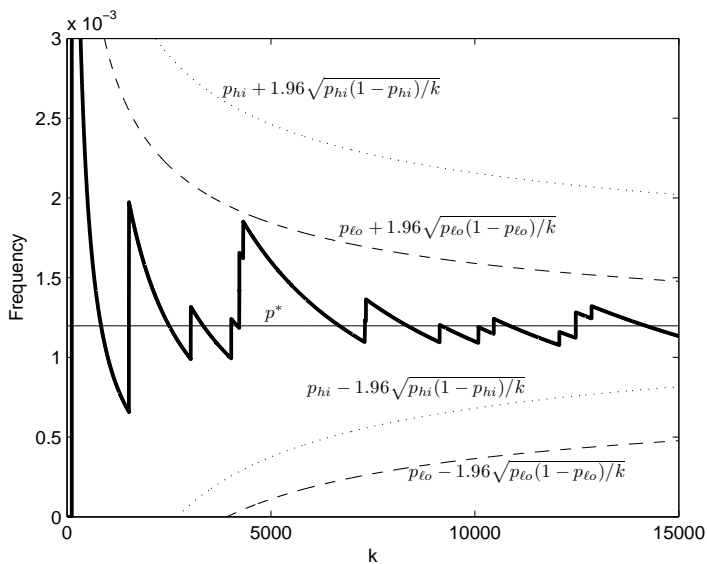}
%originally produced as eps output of SIMRWplane.m in C:\Program Files\MATLAB\R2006b\work\PROGRAM
\end{center}
\caption {Average crossing-path frequency $F(k)$ (over the first $k$ time-steps)   for  two random walks in a circle of radius $r =10$ with $d_1=1$ and $d_2=0.7$ (15,000 time-steps). 
At each time-step  the intervals within which the "hypothetical frequency" $F_h(k)$   falls  with probability $0.95$  are given for the low and high approximations  $p_{\ell o}$ and $p_{hi}$  of  $P(d_1,d_2,\rho)$.  The mid-point approximation $p^{*}$ is also plotted.}
\label{figsim}
\end{figure}

Figure \ref{figsim} depicts a simulated trajectory of the frequency $F(k)$.  We also plotted  the hypothetical low and high  intervals  within which each $F_h(k)$ would fall   with probability $0.95$. These bands shed light on the expected fluctuations of the frequency, in the case of independent events taking place with probability  $p_{\ell o}$ (or $p_{hi}$). 

This and other simulations suggest that  $p^*$  gives at least an idea of the (long-run) crossing-path probability.  

There is less uncertainty on the sphere as we have shown that the one-step probability of intersection  $P_s(d_1,d_2,R)$ is  extremely close to  $d_1d_2/(2\pi^2\rho^2)$.   Simulations performed on the sphere (not shown)  yield long-run crossing-path rates  that are close to $d_1d_2/(2\pi^2\rho^2)$ even though the law of large numbers cannot be invoked because of spatial dependence over time. 

\section{Discussion}\label{disc}
The expression of Eq. \eqref{P2200}  for the crossing-path probability  in the plane can be relatively  crude. However it has the merit of simplicity and 
 it improves  if the area of the  region $R$ increases. 

On the sphere the crossing-path probability  can be approximated very closely by the  simple expression $d_1d_2/(2\pi^2\rho^2)$. We  derived  this expression on the basis of an analytical result  for the  plane (Appendix A), expecting it to be a good  approximation only for a sphere of infinitely large radius.  It is of some interest to note that because of the complexity of the multiple integrals in spherical coordinates  we saw now way of obtaining  this  approximation   from the calculations performed directly on the sphere (Appendix C). 

Because of  marked departures from required assumptions  the law of large numbers could be applied neither in the plane nor on the sphere.  Despite that, the long-run average crossing-path probabilities appear to be close to the first-step probabilities.   This suggests that  a weaker version of the law of large numbers may be applicable. For example results on "weakly dependent" random variables (i.e. variables that  are  "m" (or "$\varphi$")-dependent (\cite{Sun04})) may provide more insights into  the long-run  behavior of the system. 

We note some obvious and some  less obvious extensions that can be of use to population biologists (and perhaps others):

\begin{itemize}
\item  If the crossing of paths takes place between an infected and a susceptible agent, 
then the transmission of the infection  may occur with only a probability $\tau$. In such a case the crossing-path probabilities found here need simply be multiplied by $\tau$ in order to obtain 
the probability of transmission at each time-step.  
\item An important extension would have $I$ such infectives and $S$ such susceptibles.   Epidemiologists would be keenly  interested 
in \textit{analytical results} on  the rate at which the infection would then  spread. 
\item The assumption that an agent moves at an angle that is uniformly distributed in (0,2$\pi$) may not be realistic.  For example animals may move only within  a limited angle in the continuation 
of the previous direction.  Preliminary investigations suggest that the results obtained here may still be applicable. 
\end{itemize}

The results given in this paper are merely  starting points for more in-depth theoretical investigations. They also provide  practitioners with some 
answers concerning the dynamics  of a process that depends on randomly moving agents meeting in a spatially explicit environment.

% comparison with eps def of meeting 
% see what happens with multiple agents - 
% \pagebreak
 \section{Appendices}\label{app}
 \subsection{Appendix  A: Proof of Proposition \ref{PROP1}}\label{ap1}
\begin{figure}
\includegraphics{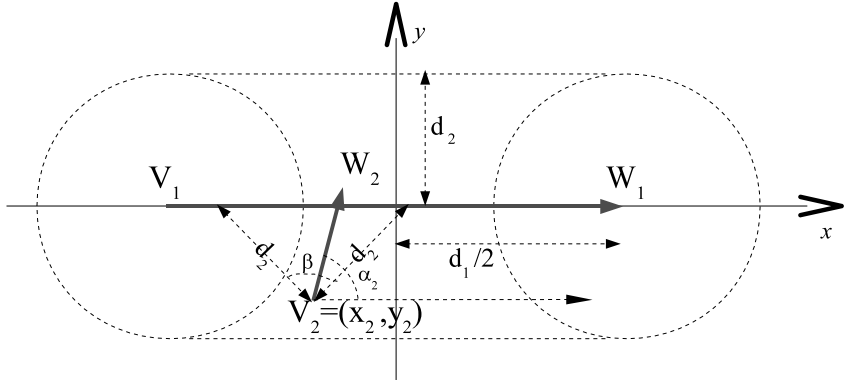} \\
\caption{Intersecting segments $V_1W_1$ and $V_2W_2$ in coordinate system in which $V_1W_1$ lies  on the $x$-axis  and the origin is at the middle of   $V_1W_1$.} 
\label{labelfig2}
\end{figure}

The double integral  $I_p(d_1,d_2)$  in Eq.  \eqref{G10} is calculated in the orthonormal coordinate system $(x,y)$ for which the segment $V_1W_1$  lies on the $x$-axis and the origin is at the middle of the segment  (Figure \ref{labelfig2}).  

In the new coordinate system, the probability $f_p(x_1,y_1,\alpha_1,x_2,y_2)$  that $V_1W_1$ and $V_2W_2$ intersect depends only on the components $(x_2,y_2)$ of $V_2$.  If this probability is denoted $F_p(x_2, y_2)$, then $I_p(d_1,d_2)$ is 

\begin{equation}\label{G1}
 I_p(d_1,d_2)  = \int_{y=-d_2}^{y=d_2}  \int_{x=-d_1/2-\sqrt{d_2^2-y^2}}^{ x=d_1/2+\sqrt{d_2^2-y^2}  } F_p(x,y)dx dy.
\end{equation}

The probability  $F_p(x_2, y_2)$   is obtained by calculating  the magnitude of the angle  $\beta$ within which $V_2W_2$ (defined by the angle $\alpha_2$) must fall and then dividing by $2 \pi$ (see Figure \ref{labelfig2}).   In the Figure  the point $V_2$  is in neither circle and the angle $\beta$ is the vertex angle in the  isosceles triangle with apex $V_2$ and two sides of length $d_2$. 
 Therefore, when  $V_2$  is in neither circle the probability $F_p(x_2, y_2)$  of an intersection is given by the function 
\begin{equation}\label{f1}
 F_{p,1}(x_2, y_2)  \overset{def.}{ = }   \dfrac{      2\arccos \left(   \dfrac{\vert y_2 \vert}{d_2} \right)  }   {2 \pi } .        
\end{equation}
Similar geometric considerations show that if the two circles overlap (i.e. $d_1/2<d_2$) then for a point $(x_2, y_2)$ in the intersection of the two circles, the probability of intersection $F(x_2, y_2)$   is given by the function 
\begin{equation}\label{f2}
 F_{p,2}(x_2, y_2)  \overset{def.}{ =}         \dfrac{      \arctan \left(   \dfrac{x_2+d_1/2}{ \vert y_2 \vert   } \right) - \arctan \left(   \dfrac{x_2-d_1/2}{ \vert y_2 \vert   } \right)                     }   {2 \pi }   .
\end{equation}

Finally, if  $(x_2, y_2)$  is in only one of the circles, then one of the vertices  of the triangle with apex $V_2$ will be the center $V_1$ or $W_1$ of that circle. The probability of intersection $F(x_2, y_2)$   is then 
\begin{equation}\label{f3}
 F_{p,3}(x_2, y_2)  \overset{def.}{ =}    \dfrac{    \arccos \left(   \dfrac{\vert y_2 \vert }{d_2} \right) - \arctan \left(   \dfrac{x_2-d_1/2}{ \vert y_2 \vert   } \right)  }   {2 \pi }   .
\end{equation}

Long but elementary calculations show that the innermost integral in \eqref{G1}, considered a function $H(y)$ of $y$,   is now equal to 
\begin{equation}\label{G2}
 H(y)  \overset{def.}{ =}   \int_{x=-d_1/2-\sqrt{d_2^2-y^2}}^{ x=d_1/2+\sqrt{d_2^2-y^2}  } F_p(x,y)dx   = \dfrac{      d_1 \times \arccos \left(   \dfrac{\vert y \vert}{d_2} \right)  }   {\pi } . 
\end{equation}
We therefore have 
\begin{equation}\label{G3}
I_p(d_1,d_2)  =   \int_{y=-d_2}^{y=d_2} H(y)   dy =    \dfrac{2d_1d_2}{\pi}
\end{equation}
which is Eq. (\ref{G19}) and yields  Eq. (\ref{GGc}).

 The inequality in  (\ref{GGa}) results from the fact that $I_p(d_1,d_2)$ is an upper bound for the double integral 
in  Eq. \eqref{G10}   when $V_1=(x_1,y_1)$ is in the border region $R_b$. 

With  $V_1$ and  $V_2$ uniformly  distributed on $R$,  the probability  of intersection  \linebreak[4]   $P_p(d_1,d_2,R)$   is now obtained by integrating  $G_p( x_1,y_1, \alpha_1 )$ 
over  $(x_1,y_1)$   in $R=R_i\bigcup R_b$ and over  $ \alpha_1$ in  $(0,2\pi)$,  and then dividing  by $2\pi A[R]$:

\[  P_p(d_1,d_2,R) = \dfrac{\iiint_{(x_1, y_1) \in R, \alpha_1 \in (0,2\pi )} G_p(x_1, y_1     , \alpha_1 ) dx_1dy_1d \alpha_1     }   {2\pi A[R]}  \]

\[    = \dfrac{\iiint_{(x_1, y_1) \in R_i, \alpha_1 \in (0,2\pi )} G_p(x_1, y_1     , \alpha_1 ) dx_1dy_1d \alpha_1     }   {2\pi A[R]} \]

\[ +   \dfrac{\iiint_{(x_1, y_1) \in R_b, \alpha_1 \in (0,2\pi )} G_p(x_1, y_1     , \alpha_1 ) dx_1dy_1d \alpha_1     }   {2\pi A[R]}     \]

% \begin{matrix}  x_1 \\ y_1\\  \end{matrix} 
\begin{equation}  \label{P20}
  = \dfrac{2d_1d_2A[R_i]}{\pi A[R]^2} +   \dfrac{\iiint_{(x_1, y_1) \in R_b, \alpha_1 \in (0,2\pi )} G_p(x_1, y_1     , \alpha_1 ) dx_1dy_1d \alpha_1     }   {2\pi A[R]}.
\end{equation}

The first term is easily calculated while the second one is a complicated  triple integral of a double integral that can only be calculated numerically. 
 The inequality in  (\ref{GGa})  implies however that 
\begin{equation}  \label{P21}
   \dfrac{2d_1d_2A[R_i]}{\pi A[R]^2}   \le    P_p(d_1,d_2,R)  \le  \dfrac{2d_1d_2(A[R_i] +A[R_b])}{\pi A[R]^2} =    \dfrac{2d_1d_2}{\pi A[R]}.
\end{equation}
which is (\ref{P2100}).  The approximation of  (\ref{P2200}) is simply the mid-point of the interval in  (\ref{P21}).  The error term of Eq. (\ref{P24}) is based on the low and high bounds of   (\ref{P21}).

 \subsection{Appendix  B: Derivation of $W_1, W_2$}\label{ap2}
In order to find each  endpoint $W_k$ ($k=1, 2$) we will first determine  a particular point  $Q_k$ at a geodesic distance $d_k$ from $V_k$. Each $W_k$ is then obtained by rotating   $Q_k$ by a uniformly distributed angle $\alpha_k$ about the  $\overrightarrow{OV_k}$ axis. 

Each  $Q_k$ is defined as having   the same azimuthal coordinate $\theta_k$ as $V_k=(\rho,\theta_k,\phi_k )$ and a zenithal coordinate $\phi$ that puts $Q_k$ at a geodesic distance $d_k$ from $V_k$. To calculate   $Q_k$   we define the function 

\begin{equation}\label{Z1}
z(\phi,d) \overset{def.}{ = }    \left\{ \begin{array}{ll}
      \phi+d/ \rho     & \mbox{if $\phi < \pi - d/\rho $};\\
    \phi-d/ \rho       & \mbox{otherwise}.\end{array}    \right.    
\end{equation}
Then 
\begin{equation}\label{Z2}
Q_k \overset{def.}{ = }  (\rho, \theta_k, z(\phi_k,d_k)). 
\end{equation}
To  obtain $W_k$ from   $Q_k$  we  define  the normalized vectors $u_k \overset{def.}{ = }  \overrightarrow{OV_k}  / \|  \overrightarrow{OV_k} \| $ 
with Cartesian coordinates  $ u_{k,x},   u_{k,y},  u_{k,z}$.  
We next define the antisymmetric matrix  

% W_k   \overset{def.}{ = } 
\begin{equation}\label{Z3}
 A_k   \overset{def.}{ = }            \left(          \begin{array}{ccc}
    0   &  -u_{k,z}    &         u_{k,y}                \\
u_{k,z}      &  0  &  -u_{k,x}   \\
 -u_{k,y}     & u_{k,x}  &  0\\  \end{array}    \right).
\end{equation}
We let   $I_3$ denote the $3 \times 3$ identity matrix. We will also need the  $Cart(\rho,\theta, \phi)$ and $Sph(x,y,z)$  
functions that transform spherical coordinates into Cartesian ones, and vice-versa, i.e. 
\begin{equation}\label{C24}
Cart(\rho,\theta, \phi))=    \rho  \left( \cos(\theta)\sin(\phi) ,\sin(\theta)\sin(\phi) , \cos(\phi)   \right) 
\end{equation}
and

\begin{equation}\label{C25}
Sph(x,y,z)=    \left(        \sqrt{x^2+y^2+z^2}    ,  \theta (x,y,z),  \arccos  \left( \frac{z}{x^2+y^2+z^2}     \right)      \right) 
%\frac{z}{x^2+y^2+z^2}  
% \acos \frac{ z }{  x^2+y^2+z^2 }   \right) 
\end{equation}
where
\begin{equation}
 \theta(x,y,z) \overset{def.}{ = }      \left\{ \begin{array}{ll}
     \arcsin\frac{y}{\sqrt{x^2+y^2}}  &   \mbox{ if   $0 \le x $}\\
  \pi - \arcsin \frac{y}{\sqrt{x^2+y^2}}   &
\mbox{otherwise}.\end{array}    \right.  
\end{equation}
 
Each $W_k$, expressed in spherical coordinates,  is now obtained by using Rodrigues' rotation formula, (\cite{Bel})  i.e. 

\begin{equation}\label{Z4}
 W_k   \overset{def.}{ = } Sph \left(      [ I_3 + A_k\sin(\alpha_k) + A_k^2(1-cos(  \alpha_k ))   ]        Cart ( Q_k)   \right) 
\end{equation} where the   ${\alpha_k}'s$ are uniformly distributed in $(0, 2\pi)$. 

 \subsection{Appendix  C: Proof of Proposition  \ref{PROP2}}\label{ap3}

\begin{figure}[h]
\begin{center}
\includegraphics[scale=0.6] {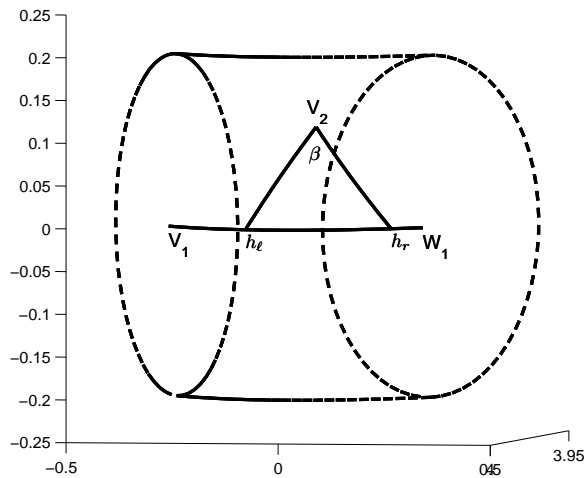}
%originally produced as eps output of grsph.m in C:\Program Files\MATLAB\R2006b\work\PROGRAM
\end{center}
\caption{Feasible domain on the sphere and angle $\beta$ within which  $V_2W_2$ must fall for the arcs   $V_1W_1$ and  $V_2W_2$ to  intersect. (When $V_2$ 
is in the right circle (resp. the left circle), $h_r$ (resp. $h_\ell$)   will be at $W_1$ (resp. $V_1$).}
\label{figde}
\end{figure}

We will derive an expression for  $I_s(d_1,d_2,\rho)$  in  the  spherical coordinate system  in which  the arc $V_1W_1$  lies on the equator and its middle is at the point $(\rho, 0, \pi/2)$,  i.e.
\begin{equation}\label{G11x}
V_1=(\rho, -d_1/(2\rho), \pi/2),  W_1=(\rho, d_1/(2\rho), \pi/2). 
\end{equation}
The feasible domain now consists of two areas on the surface of the sphere. First  the  "spherical rectangle" centered on  $(\rho, 0, \pi/2)$ with sides of geodesic lengths $d_1$ and $2d_2$; and second at both ends of the rectangle  the half-circles centered at $V_1$ and $ W_1$  and of (geodesic) radius $d_2$ (Figure \ref{figde}).  The feasible domain now depends only on $d_1, d_2$ and is denoted $FD(d_1, d_2)$. 

We need several functions in order to calculate  $ G_s(\theta_1,\phi_1, \alpha_1)$:
\begin{itemize}
\item 
The geodesic distance function between two points $P_1=(\rho,\theta_1, \phi_1)$ and $P_2= (\rho,\theta_2, \phi_2)$  on the sphere:  
 \[  gd(P_1,P_2) \overset{def.}{ =}  \]
\begin{equation}\label{G11a}
  \rho \arccos [ \sin ( \phi_1) \sin (\phi_2)( \cos ( \theta_1) \cos ( \theta_2)    + \sin ( \theta_1) \sin ( \theta_2)) + \cos ( \phi_1)\ cos ( \phi_2)].
\end{equation}
\item 
If $X$ and $Y$ are two vectors (in Cartesian coordinates) on the sphere of radius $\rho$, then the normed  tangent vector at $X$ to the geodesic line between $X$ and $Y$: 
\begin{equation}\label{G11s}
\tau (X,Y)\overset{def.}{ =} \dfrac{ Y - \frac{X^TY }{\|X\|^2}  X     }          { \| Y - \frac{X^TY }{\|X\|^2}  X     \|   }
\end{equation}
\item  The $Cart(\rho,\theta, \phi)$ function that transforms spherical coordinates into Cartesian ones (Eq. (\ref{C24})). 
\end{itemize}

Given   $V_2=(\rho,\theta_2, \phi_2)$ in the feasible domain $FD(d_1, d_2)$,  we let  $h_r$  and   $h_\ell$  be the two points on the equator (on the right and on the left of $V_2$) that determine the magnitude of the angle  $\beta$ within which  $V_2W_2$ must fall for the intersection to occur (Figure \ref{figde}).

Bearing in mind  $V_1$ and  $W_1$ of Eq. \eqref{G11x}, the Cartesian coordinates of  $h_r$  and   $h_\ell$, considered functions of $V_2=(\theta_2, \phi_2)$, are
\begin{equation*}
h_r (\theta_2, \phi_2) \overset{def.} { =} 
\end{equation*}

 \begin{equation}\label{Z100}
    \left\{ \begin{array}{ll}
    Cart(W_1)      \mbox{ if   $gd(V_2,W_1)<d_2 $}\\
    \rho          
\left(        \cos  \left(   \theta_2 +  \arccos \frac{\cos(d_2/\rho)}{ \sin(\phi_2)} \right),  \sin  \left(   \theta_2 +  \arccos \frac{\cos(d_2/\rho)}{ \sin(\phi_2)}   \right), 0    \right)    \mbox{otherwise}.\end{array}    \right.    
\end{equation}

\begin{equation*}
  h_\ell (\theta_2, \phi_2)\overset{def.} { =} 
\end{equation*}

\begin{equation}\label{Z200}
   \left\{ \begin{array}{ll}
      Cart(V_1)      \mbox{ if   $gd(V_2,V_1)<d_2 $}\\
   \rho          
\left(     \cos  \left(   \theta_2 -  \arccos \frac{\cos(d_2/\rho)}{ \sin(\phi_2)} \right),  \sin  \left(   \theta_2 -  \arccos \frac{\cos(d_2/\rho)}{ \sin(\phi_2)}   \right) ,  0   \right)    
\mbox{otherwise}.\end{array}    \right.    
\end{equation}

In the new coordinate system, the probability  of intersection  $f_s(\theta_1,\phi_1,\alpha_1,\theta_2,   \phi_2)$ conditionally  on the arc $V_1W_1$ and  on $V_2$   depends only on the components $(\rho, \theta_2,\phi_2)$ of $V_2$  and  is denoted   $F_s(\theta_2, \phi_2)$. Given the uniformity assumption for the angle  $\alpha_2$ at which the second agent leaves $V_2$ to go to $W_2$, the probability is equal to the magnitude of the angle  $\beta$ within which the arc $V_2W_2$ must fall for the intersection to occur, divided by $2\pi$.   The angle $\beta$ is the angle between  the tangent vectors at $V_2$ in the directions  of  $h_\ell (\theta_2, \phi_2)$ and  $h_r(\theta_2, \phi_2)$.    The probability of intersection   $F_s(\theta_2, \phi_2)$ is therefore

\begin{equation}\label{Z300}
F_s(\theta_2, \phi_2) = \dfrac{\arccos  \left(        \tau[Cart(\rho,\theta_2, \phi_2), h_r (\theta_2, \phi_2)]^T      \tau[Cart(\rho,\theta_2, \phi_2), h_\ell (\theta_2, \phi_2)]         \right)    }{  2\pi}          
\end{equation}

\begin{table}
\caption{Percentage error $100\times\left(  \frac{ 2d_1d_2/\pi}{I_s(d_1,d_2,\rho)} -1 \right) $  when approximating $I_s(d_1,d_2,\rho)$ as $2d_1d_2/\pi$, with $d_1=3$ and 
an illustrative  range of values for $d_2$ and $\rho$.}
\label {T1}
\begin{tabular}{|c|c|c|c|c|} 
% calc below in ProofSphere.mcd
\hline $d_2\downarrow; \rho \rightarrow$   & 2 &  3  & 4  & 5  \\ 
\hline    1   & $4.866 \times 10^{-4}$ & $-1.68 \times 10^{-4}$ &  $-2.527 \times 10^{-4}$ & $-8.471 \times 10^{-4}$    \\ 
\hline    2   & $1.288 \times 10^{-3}$ & $5.595 \times 10^{-4}$ &  $ 9.269 \times 10^{-4}$ & $6.658 \times 10^{-4}$    \\ 
\hline    3   & $-1.506 \times 10^{-5}$ & $7.529 \times 10^{-4}$ &  $ 9.412 \times 10^{-5}$ & $8.954 \times 10^{-4}$    \\ 
\hline 
\end{tabular} 
\end{table}

The double integral  $I_s(d_1, d_2,\rho)$  is  now calculated  in the new coordinate system by integrating $F_s(\theta_2, \phi_2)$ over the feasible domain $FD(d_1, d_2)$.   Because of symmetries the integral    $I_s(d_1, d_2,\rho)$  is the sum of  four times the integral  over the upper right  quarter of the rectangular area of $FD(d_1, d_2)$  and of   four times the integral  over the upper half of the right circle.  We thus  have 
\begin{equation*}
  I_s(d_1, d_2,\rho)=  4\rho^2      \int_{\theta_2=0}^{d_1/(2\rho)}      \int_{\phi_2=\pi/2-d_2/\rho}^{\pi/2}            F_s(\theta_2,\phi_2)sin(\phi_2)d\theta_2d\phi_2     + 
\end{equation*}

\begin{equation}\label{Z400}
 4\rho^2      \int_{\theta_2=d_1/(2\rho)}^{d_1/(2\rho) + d_2/\rho}      \int_{\phi_2=\phi(\theta_2)}^{\pi/2}            F_s(\theta_2,\phi_2)sin(\phi_2)d\theta_2d\phi_2         
\end{equation}
where 
\begin{equation}
\phi(\theta_2)   \overset{def.}{ =}   \arcsin            \left(      \dfrac{    \cos \frac{d_2}{\rho}  }{     \cos  \left(    \theta_2 -  \frac{d_1}{2\rho}  \right)          }                     \right) 
\end{equation}
is the lower value of $\phi_2$ when integrating in the right half circle. 

Over a wide range of values for  $d_1, d_2$ and $\rho$  we found that the relative error made when approximating    $I_s(d_1, d_2,\rho)$ of Eq. (\ref{Z400})  as  $2d_1d_2/\pi$   was of the order of $10^{-3}$  to  $10^{-5}$  \textit{percent}. See Table  \ref{T1} for an example of this relative error for a range of values of $d_2$ and $\rho$.  In fact the error is so small that it is within the margin of error when calculating $I_s(d_1, d_2,\rho)$ numerically.    The fact that  $I_s(d_1, d_2,\rho)\approxeq2d_1d_2/\pi$  combined with Eq.  (\ref{G10u})  yields the result of  (\ref{GGb}).

\end{document}